\documentclass[twocolumn]{aastex61}
\usepackage{booktabs}

\shorttitle{MONDian gravity in pressure supported systems}
\shortauthors{R. Durazo, X. Hernandez, B. Cervantes Sodi and S. F. Sanchez}

\begin{document}


\title{A Test of MONDian Gravity in $\sim$300 pressure supported elliptical galaxies from the MaNGA survey}


\author{R. Durazo$^{1}$, X. Hernandez$^{1}$, B. Cervantes Sodi$^{2}$ and S. F. Sanchez$^{}$}

\affil{$^{}$Instituto de Astronom\'{\i}a, Universidad Nacional Aut\'{o}noma de M\'{e}xico, Apartado Postal 70--264 C.P. 
04510 M\'exico D.F. M\'exico.\\
$^{2}$Instituto de Radioastronom\'{\i}a y Astrof\'{\i}sica, Universidad Nacional Autonoma de M\'exico, Campus Morelia, 58090 Morelia, 
Michoac\'an, M\'exico.\\}



\begin{abstract}
Pressure supported systems modeled under MONDian extended gravity are expected to show an outer flattening in their velocity dispersion profiles. A characteristic scaling between the amplitude of the asymptotic velocity dispersion and the radius at which the flattening occurs is also expected. By comprehensively analyzing the dynamical behavior of $\sim$300 extremely low rotating elliptical galaxies from the MaNGA survey, we show this type of pressure supported system to be consistent with MONDian expectations, for a range of central velocity dispersion values of $60 km/s < \sigma_{central}< 280 km/s $ and asymptotic velocity dispersion values of $28km/s < \sigma_{\infty}<250 km/s$. We find that a universal velocity dispersion profile accurately describes the studied systems; the predicted kinematics of extended gravity are verified for all well observed galaxies.
\end{abstract}



\section{Introduction}

In recent years, high expectations were placed on direct dark matter detection searches by the LHC (CMS collaboration 2016), PANDAX-II (Yang et al. 2016) and LUX (Szydagis et al. 2016) experiments, as well as on indirect dark matter detection by pair annihilation (Fermi-LAT and DES collaborations 2016), encountering all null detection signals to date. Even the most recent results from the XENON1T (Aprile et al. 2017) have come up empty, continuing the 30 year dry spell of the dark matter detection quest and motivating the undergoing exploration of alternative explanations to the dynamical effects appearing in low acceleration regimes.
The most  successful alternative to dark matter on the galactic phenomenological level is modified Newtonian dynamics (MOND) put forth by Milgrom (1983), in which baryonic matter distributions are the only elements required to reproduce observed discrepancies from Newtonian gravity in measured kinematics. Furthermore, recent theoretical developments indicate the possibility of the emergence of new fundamental descriptions of gravity, e.g. emergent gravity (Verlinde 2016) and covariant torsion models (Barrientos \& Mendoza 2017). 

The issue however, remains controversial, due in large part to the difficulty in fitting the CMB angular power spectrum without introducing an extra degree of freedom decoupled from the baryon-photon plasma at recombination (Slosar et al. 2005, Angus 2009). This extra degree of freedom is generally ascribed to a dark matter component e.g. (Plank Collaboration 2016).
Within the framework of modified theories of gravity, cosmological tests must necessarily be framed within extensions to GR, (e.g. Capozziello \& de Laurentis 2011, Nojiri \& Odintsov 2011) and while no single such proposal is currently close to reproducing the details of the $\Lambda$CDM CMB fit, this last model also displays worrisome tensions at cosmological scales. A recent example of the above can be found in the inconsistency between CMB $\Lambda$CDM fits to Plank data requiring a Hubble constant of $66.93 \pm 0.62$ $km s^{-1}$ Mpc$^{-1}$ and direct supernova observations of the local cosmological expansion rate yielding values of $73.24 \pm 1.74$ $km s^{-1}$ Mpc$^{-1}$, e.g. Riess et al. (2016). Tension appears also between the cosmological parameters $\Omega_m$ and $\sigma_8$ from cluster studies (B{\"o}hringer et al. 2014) and those derived from CMB anisotropies (Planck 2016), always under the $\Lambda$CDM framework. Cosmological constraints thus provide guidelines in orienting relativistic modified gravity theories seeking to be at terms with observations at the largest scales.

Still, the succes of modified gravity at galactic and sub galactic scales is clear (Sanders \& McGaugh 2002, Famaey \& McGaugh 2012), and it has been shown to better explain velocity profiles than dark matter (Swaters et al. 2010; McGaugh 2012; Lelli et al. 2017). 
The success of modified gravity is most dramatically evident for spiral galaxies (Begeman et al. 1991, Sanders 1996, de Blok \& McGaugh 1998, Gentile et al. 2011, Milgrom \& Sanders 2007; McGaugh 2016; Desmond 2017), where the rotation curves can be quite a precise tracer of the gravitational force; in hot stellar systems with little rotational support, the predictions are less straightforward. Recently, a variety of pressure supported astrophysical systems have been studied and compared to MONDian predictions, such as Galactic globular clusters (Gentile et al. 2010; Ibata et al. 2011; Sanders 2012; Hernandez \& Jimenez 2012; Hernandez et al. 2017; Thomas 2018), dwarf spheroidal galaxies (Milgrom 1995; Brada \& Milgrom 2000; McGaugh \& Wolf 2010; McGaugh \& Milgrom 2013; Lughausen et al. 2014; Alexander et al. 2017) and elliptical galaxies (Sanders 2000; Milgrom \& Sanders 2003; Tiret et al. 2007; Richtler et al. 2011; Schubert et al. 2012; Milgrom 2012; Jimenez et al. 2013).

In particular, a successful test of MONDian gravity in two types of pressure supported systems, Galactic globular clusters and elliptical galaxies, was performed in Durazo et al. (2017), in which a universal projected velocity dispersion profile was shown to accurately describe the two types of astrophysical systems, and that the expectations of MONDian gravity were reproduced across seven orders of magnitude in mass. However, due to the high scatter in velocity dispersion measurements of elliptical galaxies from the CALIFA project (Sanchez et al. 2012; Walcher et al. 2014; Sanchez et al. 2016a) resulting in large errors in the profile fits, and the fact that well measured proper motions for Galactic globular clusters are required to discard tidal heating as a viable explanation for the outer flattening of the velocity dispersion profile, the sub-sets of the two classes of objects were relatively modest, only a small sample of 13 well measured velocity dispersion profiles of elliptical galaxies were used in said study. It is therefore desirable to extend the size of the sample to strengthen the statistical analysis, which we do here by using newly released velocity dispersion measurements from the MaNGA survey (Bundy et al. 2015, Law et al. 2015, 2016, Yan et al. 2016a, 2016b, Wake et al. 2017), showing significantly less scatter and higher sensitivity to low velocity data.

As discussed in our previous research, we work with the theoretical scalings of two directly observable velocity dispersion parameters; the asymptotic value of the velocity dispersion profile, $\sigma_{\infty}$, and the characteristic radius beyond which a flattening is apparent, $R_{M}$. This eliminates the problematic process of baryonic mass determinations that carry various uncertainties and systematics such as hard to determine gas and dust fractions, star formation histories, unknown stellar mass functions, varying mass to light rations, multiple and complex stellar populations, as well as radial variations in all the aforementioned parameters. We adopt the proposed universal velocity dispersion profile used in Durazo et al. (2017), which has been shown to correctly reproduce velocity dispersion profiles in pressure supported systems.

In section 2 we construct first order equations for predicting the asymptotic velocity dispersion expected under MONDian gravity, and the corresponding relations between the parameters describing the velocity dispersion profiles for such pressure supported systems. Section 3 describes the selection of data used to build working samples of velocity dispersion profiles for elliptical galaxies, as well as the fitting methodology and estimated parameters for the universal velocity dispersion profile used. In section 4 we show that the flattening radius and asymptotic velocity dispersion follow a scaling consistent with the generic predictions of MONDian gravity. Section 5 outlines our findings.
~\\

\section{MONDian Theoretical expectations}

In order to model observed galactic dynamics assuming only baryonic matter, a change from the Newtonian force law $F_{N}=GM/r^{2}$ to a MONDian force law $F_{M}=(GM a_{0})^{1/2}/r$ is needed at a scale of $R_{M}=(G M /a_{0})^{1/2}$ e.g. Milgrom (1983).
Most authors seem to agree on a relatively abrupt transition between the above regimes, from studies on Milky Way rotation curve comparisons e.g. Famaey \& Binney (2005), solar system dynamics e.g. Mendoza et al. (2011).

Centrifugal equilibrium velocities will show a Tully-Fisher value in the MONDian regime of:

\begin{equation}
V=\left( G M a_{0}\right)^{1/4},
\end{equation}

\noindent for a test particle orbiting a total baryonic mass M. 
Very generally for MONDian theories, for any isolated self-gravitating system in the deep-MOND regime (beyond $R_M$), a relation of the form :

\begin{equation}
\alpha MGa0=v^4,
\end{equation}

\noindent will hold, where $v$ is a characteristic measure of the mean 3D velocity, and $\alpha$ is a dimensionless parameter which depends on the exact theory being considered and on the details of the system, such as the slope of its mass profile. For example, the deep MOND virial relation of the Bekenstein-Milgrom theory where $v$ becomes the 3D isotropic velocity dispersion of a system and $\alpha= 4/9$ (see McGaugh \& Milgrom 2013; Famaey et al. 2018). We remain within MONDian limits for isolated systems, as our selection criteria define a sample where any external field effect should be minimum (see section 3), and seeking a simple and generic correspondence to the inferred projected velocity dispersion of an observed system independent of any particular MOND theory, we follow McGaugh \& Wolf (2010) and assume $v=\sigma$, $\alpha=1/9$, yielding:

\begin{equation}
\sigma_{\infty}^{2}= \frac{1}{3} \left( G M a_{0}\right)^{1/2}.
\end{equation}

To test equation (3), one would need independent estimations of the baryonic mass of each system as well as observations of $\sigma_{\infty}$. Correctly estimating baryonic masses for any astrophysical system is a complicated endeavor hampered by gas and dust fractions, SFH and IMF suppositions, the complex stellar populations of elliptical galaxies, different mass-to-light ratios for varying wavelengths, as well as potential radial changes of the aforementioned parameters. Simply substituting $R_{M}$ into equation (3), one obtains a straightforward test:

\begin{equation}
R_{M}=\frac{3 \sigma_{\infty}^{2}}{a_{0}},
\end{equation}

\noindent which in astrophysical units results in:

\begin{equation}
\left( \frac{R_{M}}{pc}\right) = 0.81 \left( \frac{\sigma_{\infty}}{km/s} \right)^{2}.
\end{equation}

Equation (5) now links two directly observable features of the velocity dispersion profile of a pressure supported system, where $\sigma_{\infty}$ is the asymptotic projected velocity dispersion at large radii and $R_{M}$ is the characteristic radius at which the velocity dispersion profile begins to flatten. For equation (5) to have physical meaning, an inner Newtonian region with a Keplerian decline in the velocity dispersion profile is required, as well as a clear transition to a flat velocity dispersion regime after reaching scales of $R_{M}$. 
Recently in Durazo et al. (2017), observed velocity dispersion profiles of 13 extremely low rotation ellipticals from the CALIFA survey (Sanchez et al. 2016b). Results indicate a strong compliance with MONDian expectations, inspiring us to build on these findings and perform a similar study for a much larger sample of extremely low rotating elliptical galaxies from the MaNGA survey (Bundy et al. 2015, Law et al. 2015, 2016, Yan et al. 2016a, 2016b), which forms part of the larger SDSS-IV project (Gunn et al. 2006, Smee et al. 2013, Blanton et al. 2017).

In the following section we show almost 300 velocity dispersion profiles for our sample of low rotating ellipticals having the previously discussed behavior, which in section 4 we compare to the expectations of equation (5).
Reminiscent of the purely Newtonian expression correlating half mass-radius and central velocity dispersion (Cappellari et al. 2006; Wolf et al. 2010) is the expression $R_{M}\/\sigma_{\infty}^2=GM/3$ resulting from multiplying equations (3) and (4) and substituting $M$ for $\sigma_{\infty}$. 

Here we have assumed that the total baryonic mass of the system has converged in the regime where the velocity dispersion profile flattens, allowing for the treatment of the galaxy in question as a point mass, in accordance with Newton's theorems for spherically symmetric mass systems valid under MONDian gravity (Mendoza et al. 2011). This allows for the treatment of the dynamical tracers from which $\sigma_{\infty}$ is estimated, as test particles. Given the approximate de Vaucouleurs inferred surface brightness profiles of elliptical galaxies (although most elliptical galaxies do not strictly follow an $n=4$ Sersic profile, they are generically characterized by very centrally concentrated light profiles), this will in general be a good approximation. 

Recently, pressure supported systems have been successfully reproduced under MONDian gravity (e.g. Gentile 2010; Sanders 2010; Hernandez et al. 2013b) and in particular, Jimenez et al. (2013) accurately modeled the giant elliptical galaxy NGC4649, Chae \& Gong (2015) reproduced observed velocity dispersion profiles using MONDian models requiring only stellar mass, and Tian \& Ko (2016) found that the dynamics of seven elliptical galaxies traced by planetary nebulae were properly explained by MOND. However, Richtler et al. (2008) and Samurovi{\'c} (2014, 2016) found a dark matter halo to be a better fit than MOND in elliptical galaxies, thus the item remains a matter of debate.

\section{Empirical velocity dispersion profiles}
In Durazo et al. (2017), a sub sample of extremely low rotation galaxies was extracted from the second data release of the CALIFA survey (S{\'a}nchez et al. 2012; Walcher et al. 2014), in which 200 radial 2D velocity dispersion profiles were produced using the Pipe 3D pipeline (S{\'a}nchez et al. 2016b). The sub sample studied comprised galaxies with an average value of maximum rotation velocity to central velocity dispersion per galaxy of $V_{max}/\sigma_0=0.213$, and negligible gas content. In Durazo et al. (2017) we showed that a universal velocity dispersion profile of the form:

\begin{equation}
\sigma(R)= \sigma_{0} e^{-(R/R_{\sigma})^{2}} + \sigma_{\infty}
\end{equation}

\noindent accurately reproduces the reported velocity dispersion profiles. In the above equation, $\sigma_{\infty}$ is the velocity dispersion asymptotic flattening value, the central velocity dispersion of the system is given by $\sigma(0)=\sigma_{0}+\sigma_{\infty}$, and $R_{\sigma}$ represents the radius after which the value of $\sigma_{\infty}$  is quickly approximated.

Building on the successful fitting of equation (6) for the 13 low rotation CALIFA galaxies, we construct here a considerably larger sample of such systems and perform equivalent fits to our universal velocity dispersion equation, as well as a test of the MONDian prediction of eq. (5) to our data.

Unlike previous SDSS surveys which obtained spectra only at the centers of target galaxies, MaNGA (Bundy et al. 2015, Law et al. 2015, 2016, Yan et al. 2016a, 2016b) provides spectral measurements across the face of $\sim$10,000 nearby galaxies thanks to 17 simultaneous “integral field units” (IFUs), each composed of tightly-packed arrays of optical fibers (Drory et al. 2015, Law et al. 2015). From nearly 3000 elliptical galaxies analyzed using Pipe 3D pipeline (Law et al. 2016, Sanchez et al. 2017a), only galaxies with $V/\sigma<0.213$ were selected, in accordance with the values used as selection criterion in Durazo et al. (2017), to ensure that our selection contains only systems with minimal dynamical support besides velocity dispersion. 

Following the methodology used in our previous work, we average the observed velocity dispersion data in 20 radial bins each containing the same number of data points (Figure 1), with a corresponding propagated average velocity dispersion error. All data points below the threshold of 20 $km\/s^{-1}$ are discarded as they are considered to be unreliable (Bundy et al. 2015, Yan et al. 2016b). The requirement of fitting each whole galaxy within the MaNGA IFU instrument implies that the more massive galaxies will necessarily be more distant, and hence systematically less accurately observed.
We use a nonlinear least squares Levenberg-Marquardt algorithm to fit the universal function of equation (6) to the projected velocity dispersion profiles and estimate the equation parameters along with their respective confidence intervals. We discard systems with poorly estimated parameters and unclear morphology, as well as fits with fractional errors larger than 50\% of the estimated parameters. Galaxies not covered to at least $2R_e$ were not considered, to ensure that we are indeed reaching the outer region of each galaxy. We thus obtain a sub sample composed of 292 low rotating well fitted ellipticals (Table 1). 

\onecolumngrid
\begin{longtable}{llccrrccrr}
\caption{Fitted parameters with respective errors, rotation velocity to velocity dispersion ratio measured at half-light radius, and stellar masses estimated from Sanchez et al. (2017a) for our 292 ellipticals.}\\
\toprule
ID & MaNGA name &  $\sigma_0$ &  $\delta\sigma_0$ &  $R_{\sigma}$ &  $\delta R_{\sigma}$ &  $\sigma_{\infty}$ &  $\delta\sigma_{\infty}$ &  $V/\sigma$ &  log $M_{\star}$ \\
{} & {} & $[km/s]$ & $[km/s]$ & $[kpc]$ & $[kpc]$ & $[km/s]$ & $[km/s]$ & {}& $[M_{\odot}]$ \\
\midrule
\endhead
1   &   manga-8274-6103 &    35.9 &       11.6 &     5.4 &        2.1 &     214.1 &          7.7 &   0.039 &     11.1 \\
2   &   manga-8083-1902 &    54.2 &       10.4 &     2.8 &        0.7 &     151.9 &          7.8 &   0.039 &     10.7 \\
3   &   manga-8256-6103 &    36.4 &       11.7 &     6.8 &        2.8 &     212.9 &         10.4 &   0.040 &     11.2 \\
4   &   manga-8451-6102 &    24.7 &       12.4 &     7.0 &        4.3 &     231.9 &         11.3 &   0.041 &     11.2 \\
5   &   manga-8553-3702 &    68.0 &        9.1 &     3.9 &        0.7 &     115.1 &          6.8 &   0.042 &     10.9 \\
6   &   manga-8604-6102 &    53.4 &       10.6 &     2.7 &        0.7 &     163.6 &          6.4 &   0.042 &     10.8 \\
7   &   manga-8333-6104 &    28.9 &       11.9 &     6.4 &        2.7 &     207.0 &          5.9 &   0.042 &     11.3 \\
8   &   manga-8319-6101 &    46.4 &       11.1 &     5.2 &        1.4 &     188.0 &          7.6 &   0.044 &     11.1 \\
9   &   manga-8567-6104 &    17.1 &       11.6 &     2.6 &        1.8 &     142.3 &          4.9 &   0.045 &     10.9 \\
10  &   manga-8549-6101 &    38.5 &       11.5 &     4.3 &        1.6 &     173.1 &          7.3 &   0.048 &     11.1 \\
11  &   manga-8566-6104 &    50.5 &       24.1 &     5.4 &        3.2 &     138.8 &         28.1 &   0.049 &     11.1 \\
12  &  manga-8249-12705 &    61.3 &       24.8 &     5.0 &        2.5 &     151.3 &         28.9 &   0.051 &     10.8 \\
13  &   manga-8247-6101 &    58.7 &       12.7 &     1.6 &        0.4 &     148.1 &          5.4 &   0.051 &     10.7 \\
14  &   manga-8978-3702 &    43.1 &       10.5 &     4.3 &        1.4 &     132.8 &          9.5 &   0.051 &     11.0 \\
15  &   manga-8588-3704 &    57.4 &       10.3 &     3.6 &        0.8 &      38.9 &         11.6 &   0.053 &     10.3 \\
16  &   manga-8253-1901 &    65.9 &        8.5 &     1.7 &        0.3 &      68.4 &          7.1 &   0.053 &     10.3 \\
17  &   manga-8253-6103 &    70.0 &       11.1 &    10.7 &        2.2 &     173.4 &          9.1 &   0.055 &     11.3 \\
18  &  manga-8315-12702 &    42.3 &       11.3 &     8.7 &        2.8 &     179.8 &          9.2 &   0.056 &     11.2 \\
19  &   manga-8440-6104 &    38.1 &       30.6 &     5.5 &        2.3 &      92.9 &         33.9 &   0.058 &     10.6 \\
20  &   manga-8459-3701 &    34.5 &       12.8 &     3.1 &        1.1 &     171.0 &          4.7 &   0.059 &     10.9 \\
21  &  manga-8952-12704 &    34.8 &       12.6 &     3.6 &        1.3 &     179.4 &          4.8 &   0.062 &     11.1 \\
22  &   manga-8338-3703 &    46.7 &        8.9 &     3.7 &        1.0 &      85.0 &          9.1 &   0.063 &     10.5 \\
23  &   manga-8239-3703 &    69.7 &       42.9 &     5.3 &        3.0 &     108.0 &         46.5 &   0.063 &     10.7 \\
24  &   manga-8485-9101 &    43.5 &       11.3 &     6.0 &        1.8 &     196.7 &          6.6 &   0.063 &     11.3 \\
25  &   manga-7975-6103 &    67.2 &        8.4 &     7.6 &        1.4 &     118.8 &          7.2 &   0.063 &     11.1 \\
26  &   manga-8155-3702 &    32.2 &       12.7 &     3.9 &        1.8 &      90.6 &         13.8 &   0.064 &     10.4 \\
27  &   manga-8947-1902 &    26.0 &        7.8 &     1.6 &        0.6 &      88.0 &          5.6 &   0.065 &     10.3 \\
28  &  manga-8724-12703 &    93.6 &       10.8 &     5.3 &        0.9 &     176.5 &          8.2 &   0.065 &     11.0 \\
29  &   manga-8604-6103 &    44.2 &       10.3 &     3.5 &        1.1 &     163.8 &          7.7 &   0.066 &     10.9 \\
30  &   manga-8456-3702 &    78.7 &       10.5 &     5.3 &        0.9 &     144.0 &          9.0 &   0.066 &     10.9 \\
31  &   manga-8319-6104 &    24.9 &       11.0 &     3.5 &        1.7 &     191.7 &          6.4 &   0.066 &     10.9 \\
32  &   manga-8550-6101 &    39.2 &        9.7 &     3.2 &        1.1 &     150.0 &          7.7 &   0.066 &     10.7 \\
33  &   manga-8333-9101 &    38.0 &       10.2 &     6.8 &        2.2 &     178.6 &          7.5 &   0.067 &     11.2 \\
34  &   manga-8453-6103 &    67.4 &       10.0 &     5.7 &        1.0 &     163.0 &          7.7 &   0.068 &     10.9 \\
35  &   manga-8553-6102 &    49.0 &        9.9 &     4.1 &        1.2 &     159.3 &          7.4 &   0.069 &     11.1 \\
36  &   manga-8138-3703 &    54.5 &       10.6 &     4.2 &        1.1 &     165.2 &          8.2 &   0.069 &     11.1 \\
37  &   manga-7443-3702 &    60.5 &        9.1 &     7.3 &        1.4 &     130.9 &          6.6 &   0.069 &     11.0 \\
38  &   manga-8728-3703 &    57.1 &       11.9 &    13.5 &        3.8 &     179.2 &         12.0 &   0.070 &     11.5 \\
39  &   manga-8459-6104 &    60.3 &        9.9 &     5.2 &        1.0 &     153.1 &          6.5 &   0.070 &     11.1 \\
40  &  manga-10001-6104 &    80.7 &       10.5 &     7.4 &        1.2 &     165.0 &          7.8 &   0.070 &     11.2 \\
41  &   manga-8612-6103 &    79.9 &       10.9 &     5.9 &        1.0 &     177.7 &          7.6 &   0.071 &     11.1 \\
42  &   manga-8239-6103 &    92.4 &       11.7 &     5.5 &        0.7 &     166.0 &          4.8 &   0.071 &     11.2 \\
43  &   manga-8255-6101 &    38.0 &       11.1 &     6.9 &        2.8 &     133.9 &         11.7 &   0.071 &     10.9 \\
44  &   manga-8252-1902 &    60.6 &        8.2 &     0.9 &        0.1 &      35.4 &          5.1 &   0.071 &      9.6 \\
45  &   manga-8143-3704 &    64.2 &       12.5 &     2.4 &        0.5 &     200.2 &          6.9 &   0.072 &     10.9 \\
46  &   manga-8452-6103 &    28.9 &       14.3 &     3.4 &        1.6 &     188.6 &          4.5 &   0.072 &     11.1 \\
47  &   manga-8555-6104 &    73.8 &       12.1 &     3.6 &        0.6 &     167.2 &          5.7 &   0.072 &     11.1 \\
48  &   manga-8484-9102 &    32.8 &       11.7 &     2.9 &        1.0 &     146.6 &          4.2 &   0.074 &     10.9 \\
49  &   manga-8931-6103 &    53.3 &        7.7 &     2.8 &        0.5 &      70.1 &          7.0 &   0.074 &     10.3 \\
50  &   manga-8726-3702 &    57.7 &        7.2 &     3.8 &        0.7 &      76.5 &          6.1 &   0.074 &     10.4 \\
51  &   manga-8613-3704 &    49.9 &       10.7 &     3.6 &        1.1 &     159.6 &         10.3 &   0.075 &     10.8 \\
52  &  manga-8602-12704 &    80.1 &       10.6 &     4.9 &        0.7 &     142.4 &          5.1 &   0.075 &     11.2 \\
53  &   manga-8551-6103 &    30.7 &        9.8 &     6.2 &        2.5 &     174.7 &          7.9 &   0.075 &     11.0 \\
54  &   manga-8466-6104 &    46.1 &        9.7 &     8.3 &        2.0 &     157.4 &          6.2 &   0.076 &     11.2 \\
55  &   manga-9042-1901 &    46.5 &       12.4 &     5.4 &        2.1 &     157.0 &         13.8 &   0.076 &     10.6 \\
56  &   manga-8261-6101 &    53.4 &       11.3 &     7.2 &        2.1 &     196.8 &          9.7 &   0.076 &     11.2 \\
57  &   manga-8156-3704 &    47.5 &       12.2 &     9.3 &        2.9 &     213.7 &         11.0 &   0.077 &     11.2 \\
58  &   manga-8329-6101 &    74.5 &        8.9 &     5.4 &        0.8 &     120.2 &          6.4 &   0.077 &     10.9 \\
59  &   manga-8946-3701 &    77.2 &        6.7 &     3.9 &        0.4 &      33.6 &          5.1 &   0.078 &     10.6 \\
60  &   manga-8725-3702 &    45.8 &       10.2 &     5.1 &        1.5 &     157.3 &          7.7 &   0.078 &     11.0 \\
61  &   manga-8945-6104 &    85.0 &        8.1 &     3.7 &        0.5 &      73.8 &          5.9 &   0.078 &     10.7 \\
62  &   manga-8325-6101 &    57.0 &        9.4 &     6.0 &        1.6 &     134.1 &          9.6 &   0.080 &     10.8 \\
63  &   manga-8313-3701 &    57.3 &        9.7 &     5.1 &        1.1 &     144.3 &          7.6 &   0.080 &     11.0 \\
64  &   manga-8325-6103 &    48.3 &        7.5 &     5.4 &        1.2 &      83.5 &          7.0 &   0.080 &     10.7 \\
65  &   manga-8713-3703 &    62.7 &        5.3 &     3.1 &        0.3 &      27.9 &          2.4 &   0.081 &     10.0 \\
66  &  manga-7990-12705 &    74.7 &       10.7 &     2.9 &        0.4 &     118.9 &          4.3 &   0.083 &     10.7 \\
67  &   manga-8146-3704 &    40.8 &       10.6 &     3.1 &        0.9 &     152.7 &          5.7 &   0.083 &     10.8 \\
68  &   manga-8601-9102 &    74.0 &       11.1 &     5.2 &        1.0 &     176.1 &          8.6 &   0.084 &     11.1 \\
69  &   manga-8313-3702 &    72.0 &       11.6 &     5.6 &        1.3 &     150.7 &         12.1 &   0.085 &     11.1 \\
70  &  manga-8084-12702 &    79.3 &        7.5 &     5.3 &        0.7 &      69.9 &          6.3 &   0.085 &     11.0 \\
71  &  manga-8728-12703 &    70.5 &       11.4 &     5.3 &        1.1 &     200.5 &          7.2 &   0.086 &     11.2 \\
72  &   manga-8315-3702 &    46.3 &        9.3 &     5.7 &        1.5 &      99.7 &          9.2 &   0.086 &     10.8 \\
73  &   manga-8313-3704 &    57.9 &       19.1 &     7.9 &        3.0 &     166.6 &         21.1 &   0.087 &     11.0 \\
74  &   manga-8482-6103 &    46.0 &       11.8 &     5.9 &        1.8 &     193.3 &         10.4 &   0.088 &     10.9 \\
75  &   manga-8440-6103 &    63.5 &       10.9 &     9.0 &        2.0 &     191.1 &          9.6 &   0.090 &     11.2 \\
76  &  manga-8980-12703 &    61.7 &        7.7 &     4.5 &        0.6 &      75.8 &          5.4 &   0.090 &     10.6 \\
77  &   manga-8591-6102 &    59.1 &       12.0 &     6.4 &        1.7 &     208.1 &         10.4 &   0.090 &     11.2 \\
78  &   manga-8482-6102 &    78.3 &       10.6 &     6.6 &        1.1 &     149.6 &          8.9 &   0.090 &     11.0 \\
79  &   manga-8720-6101 &    47.8 &        8.8 &     3.6 &        0.7 &     116.0 &          5.2 &   0.091 &     10.8 \\
80  &   manga-8131-1902 &    36.5 &        9.8 &     6.6 &        2.3 &     152.9 &          8.3 &   0.091 &     11.1 \\
81  &   manga-8450-3702 &    51.8 &       10.0 &     6.8 &        1.8 &     141.5 &          9.5 &   0.091 &     10.8 \\
82  &   manga-8980-3701 &    41.0 &        6.9 &     1.6 &        0.4 &      61.5 &          6.5 &   0.092 &      9.9 \\
83  &   manga-8601-1902 &    52.5 &        9.0 &     2.1 &        0.5 &     110.8 &          6.1 &   0.092 &     10.4 \\
84  &   manga-8453-3703 &    24.2 &        7.7 &     2.9 &        1.3 &      65.0 &          8.3 &   0.092 &      9.9 \\
85  &   manga-8440-3703 &    61.2 &        6.1 &     2.1 &        0.2 &      31.2 &          3.2 &   0.092 &     10.3 \\
86  &   manga-8330-3701 &    53.0 &       11.1 &     3.2 &        0.7 &     156.0 &          5.6 &   0.095 &     10.9 \\
87  &   manga-8721-1901 &    71.2 &       10.1 &     2.2 &        0.3 &     127.2 &          5.7 &   0.095 &     10.3 \\
88  &  manga-8330-12705 &    57.5 &       10.5 &     7.0 &        1.4 &     152.5 &          5.3 &   0.096 &     11.2 \\
89  &   manga-8551-3703 &    46.0 &       10.1 &     3.5 &        1.1 &     149.2 &          8.7 &   0.096 &     10.7 \\
90  &   manga-8253-6102 &    53.8 &        9.9 &     7.6 &        1.7 &     171.3 &          6.2 &   0.097 &     11.0 \\
91  &   manga-8604-3701 &    33.8 &        7.6 &     2.5 &        0.8 &      84.0 &          6.5 &   0.097 &     10.4 \\
92  &   manga-8147-1901 &    73.9 &       11.2 &     2.3 &        0.5 &      34.7 &         13.2 &   0.098 &     10.2 \\
93  &   manga-8332-3701 &    58.0 &        6.2 &     5.2 &        0.7 &      42.0 &          4.7 &   0.098 &     10.7 \\
94  &   manga-8462-6102 &    57.3 &        8.9 &     7.2 &        1.5 &     144.0 &          6.0 &   0.099 &     11.0 \\
95  &   manga-8555-9101 &    80.2 &       12.9 &     3.3 &        0.6 &     190.7 &          5.4 &   0.099 &     11.1 \\
96  &  manga-8319-12703 &    33.6 &       15.0 &     1.8 &        0.7 &     162.9 &          3.7 &   0.102 &     10.6 \\
97  &   manga-8138-3702 &   114.1 &        8.0 &     4.8 &        0.4 &      63.9 &          5.9 &   0.102 &     10.8 \\
98  &   manga-8141-6102 &    78.0 &        9.5 &     2.8 &        0.4 &     107.5 &          5.4 &   0.103 &     10.8 \\
99  &   manga-8451-6104 &    59.0 &       10.2 &     3.7 &        0.7 &     140.2 &          5.2 &   0.103 &     10.8 \\
100 &   manga-8449-3704 &    36.6 &       12.5 &     1.4 &        0.4 &     129.4 &          3.7 &   0.104 &     10.3 \\
101 &   manga-8255-6104 &    80.4 &       11.0 &     5.7 &        0.9 &     173.1 &          8.2 &   0.104 &     11.3 \\
102 &   manga-8156-6101 &    63.5 &        9.7 &     4.6 &        0.9 &     127.7 &          6.6 &   0.105 &     11.0 \\
103 &  manga-8261-12701 &    51.4 &        9.3 &     4.7 &        0.9 &     122.6 &          5.3 &   0.105 &     10.9 \\
104 &   manga-8333-6101 &    48.1 &       11.8 &     6.5 &        1.8 &     213.3 &          7.4 &   0.106 &     11.3 \\
105 &   manga-8548-6101 &    28.3 &        9.7 &     4.2 &        1.5 &     127.9 &          5.0 &   0.106 &     10.7 \\
106 &   manga-8317-3704 &    31.0 &        9.1 &     4.2 &        2.0 &      99.8 &         10.1 &   0.107 &     10.2 \\
107 &   manga-7815-3703 &    54.5 &       10.5 &     5.6 &        1.3 &     160.0 &          7.9 &   0.107 &     11.1 \\
108 &   manga-8448-9102 &    70.8 &        8.7 &     4.7 &        0.5 &      73.0 &          3.5 &   0.107 &     10.9 \\
109 &   manga-8601-1901 &    53.5 &       12.2 &     1.9 &        0.4 &     151.8 &          5.1 &   0.108 &     10.5 \\
110 &   manga-8141-6103 &    66.0 &       14.6 &     1.5 &        0.3 &     165.8 &          4.3 &   0.108 &     10.6 \\
111 &   manga-8317-3702 &    47.6 &       10.6 &     3.7 &        1.1 &     171.5 &          8.2 &   0.109 &     10.8 \\
112 &  manga-7495-12705 &    67.7 &        6.1 &     5.5 &        0.6 &      46.3 &          2.7 &   0.109 &     10.9 \\
113 &   manga-8078-6102 &    30.4 &       10.8 &     2.1 &        0.8 &     126.1 &          3.9 &   0.110 &     10.7 \\
114 &   manga-7495-1901 &    57.2 &        7.5 &     1.8 &        0.3 &      72.9 &          5.8 &   0.112 &     10.0 \\
115 &   manga-8084-1902 &    45.4 &        7.8 &     1.0 &        0.2 &      43.9 &          2.6 &   0.112 &      9.6 \\
116 &   manga-8615-3702 &    50.2 &       12.3 &     4.5 &        1.4 &     227.1 &          8.1 &   0.112 &     11.1 \\
117 &   manga-8611-6101 &    80.9 &       10.3 &     2.8 &        0.5 &     147.6 &          6.9 &   0.113 &     10.7 \\
118 &   manga-8554-6102 &    48.4 &       12.7 &     9.3 &        2.5 &     231.1 &          7.0 &   0.113 &     11.4 \\
119 &   manga-8263-3702 &    43.8 &        9.6 &     8.5 &        2.4 &     168.6 &          7.3 &   0.113 &     11.1 \\
120 &   manga-8623-6103 &    60.5 &       10.1 &     4.5 &        0.8 &     144.3 &          5.9 &   0.114 &     11.1 \\
121 &   manga-8132-9102 &    49.5 &       12.8 &     4.2 &        1.1 &     191.1 &          5.1 &   0.114 &     11.2 \\
122 &   manga-8455-6104 &    46.5 &       10.1 &     6.8 &        1.7 &     177.9 &          6.6 &   0.115 &     11.0 \\
123 &  manga-8149-12704 &    49.6 &        9.8 &     2.2 &        0.5 &     112.2 &          4.1 &   0.116 &     10.8 \\
124 &   manga-8250-9101 &    51.8 &       11.7 &     3.2 &        0.8 &     162.8 &          5.3 &   0.117 &     11.0 \\
125 &   manga-8612-3703 &    63.6 &       10.2 &     5.4 &        1.1 &     155.5 &          8.8 &   0.117 &     10.9 \\
126 &   manga-8726-1901 &    55.0 &       11.6 &     3.0 &        0.9 &     178.1 &         10.5 &   0.118 &     10.9 \\
127 &   manga-8249-6104 &    38.6 &       13.4 &     2.7 &        0.9 &     180.9 &          4.9 &   0.118 &     10.8 \\
128 &   manga-8601-6104 &    68.2 &       10.5 &     4.0 &        0.9 &     166.3 &          9.0 &   0.118 &     10.9 \\
129 &   manga-8312-3701 &   110.5 &        8.8 &     2.1 &        0.2 &      75.2 &          4.9 &   0.119 &     10.4 \\
130 &   manga-8603-1901 &    59.1 &        8.0 &     1.9 &        0.3 &      74.3 &          7.1 &   0.120 &     10.1 \\
131 &   manga-8716-3703 &    70.7 &       10.4 &     3.6 &        0.6 &     118.0 &          4.7 &   0.120 &     10.9 \\
132 &   manga-8131-3703 &    39.6 &       12.2 &     9.0 &        3.6 &     228.6 &         11.4 &   0.120 &     11.4 \\
133 &  manga-8603-12702 &    66.6 &        5.8 &     3.0 &        0.3 &      30.3 &          3.5 &   0.120 &     10.4 \\
134 &   manga-8952-6102 &    47.8 &       10.6 &     2.7 &        0.8 &     170.6 &          7.2 &   0.120 &     10.8 \\
135 &   manga-8326-3703 &    63.7 &       10.8 &     4.8 &        0.9 &     166.1 &          7.3 &   0.121 &     10.9 \\
136 &   manga-7975-3703 &    76.6 &        8.7 &     7.5 &        1.2 &     106.6 &          7.7 &   0.123 &     11.0 \\
137 &   manga-8717-3701 &    95.7 &       14.3 &     1.4 &        0.2 &     124.3 &          3.6 &   0.123 &     10.6 \\
138 &  manga-9029-12703 &    52.3 &       12.1 &     4.9 &        1.2 &     167.7 &          4.8 &   0.124 &     11.2 \\
139 &   manga-8254-1901 &    37.3 &        7.6 &     1.4 &        0.4 &      70.8 &          7.5 &   0.124 &      9.4 \\
140 &  manga-8447-12704 &    37.7 &       10.6 &     8.5 &        2.9 &     192.6 &          8.9 &   0.124 &     11.1 \\
141 &  manga-8725-12703 &    50.7 &        6.7 &     3.9 &        0.7 &      67.5 &          5.0 &   0.125 &     10.5 \\
142 &   manga-8156-3702 &   117.6 &        7.6 &     4.4 &        0.4 &      35.8 &          8.2 &   0.126 &      9.9 \\
143 &   manga-8606-6103 &    88.7 &        8.9 &     3.5 &        0.3 &      70.2 &          3.5 &   0.127 &     10.8 \\
144 &   manga-8616-6103 &    40.7 &       14.2 &     4.4 &        1.6 &     237.6 &          5.7 &   0.127 &     11.2 \\
145 &   manga-8554-3703 &    61.9 &        7.5 &     2.1 &        0.3 &      64.2 &          5.1 &   0.128 &     10.4 \\
146 &   manga-8458-6104 &    36.5 &        7.1 &     2.4 &        0.6 &      74.9 &          5.7 &   0.129 &     10.3 \\
147 &   manga-8312-3704 &    42.1 &        8.9 &     2.2 &        0.6 &     107.1 &          5.1 &   0.131 &     10.3 \\
148 &   manga-9029-3703 &    58.1 &       13.4 &     4.3 &        1.3 &     149.4 &         14.8 &   0.131 &     10.7 \\
149 &  manga-8143-12701 &    40.4 &       10.9 &     5.0 &        1.5 &     168.3 &          5.5 &   0.132 &     11.2 \\
150 &   manga-7960-6104 &    46.7 &        7.9 &     2.3 &        0.5 &      91.1 &          5.4 &   0.133 &     10.5 \\
151 &  manga-8317-12702 &    75.2 &        9.2 &    15.4 &        2.6 &      84.4 &          9.4 &   0.133 &     11.0 \\
152 &   manga-8459-1901 &    56.4 &       20.9 &     2.0 &        0.8 &      76.3 &         23.2 &   0.133 &      9.7 \\
153 &   manga-8591-6101 &    80.5 &       10.2 &     5.3 &        0.7 &     143.0 &          5.7 &   0.133 &     11.1 \\
154 &  manga-8341-12702 &    60.8 &       10.5 &     5.7 &        1.2 &     172.6 &          6.3 &   0.134 &     11.3 \\
155 &   manga-8439-1902 &    29.1 &        9.0 &     1.1 &        0.3 &      87.0 &          4.2 &   0.134 &     10.0 \\
156 &   manga-8465-1901 &    33.5 &       10.1 &     2.0 &        0.9 &     147.3 &          9.7 &   0.134 &     10.4 \\
157 &  manga-10001-3704 &    36.0 &       10.8 &     3.3 &        1.3 &      83.7 &         12.1 &   0.134 &     10.5 \\
158 &   manga-8551-6102 &    46.1 &        8.1 &     5.3 &        1.1 &     112.1 &          5.5 &   0.135 &     10.8 \\
159 &   manga-7962-1902 &    86.5 &        8.1 &     2.8 &        0.4 &      53.8 &          8.7 &   0.135 &     10.0 \\
160 &   manga-8133-3702 &    62.7 &       11.6 &     7.0 &        1.7 &     180.9 &         10.4 &   0.136 &     11.2 \\
161 &   manga-8952-3704 &    43.0 &        9.5 &     2.6 &        0.8 &      97.2 &          9.8 &   0.137 &     10.4 \\
162 &   manga-8259-3703 &    37.3 &       16.9 &     3.1 &        1.3 &     225.2 &          5.1 &   0.138 &     11.0 \\
163 &   manga-8335-1902 &    44.9 &        6.2 &     1.5 &        0.3 &      38.4 &          5.2 &   0.138 &      9.7 \\
164 &  manga-8948-12702 &    96.9 &        7.2 &     3.2 &        0.3 &      41.4 &          3.8 &   0.139 &     10.7 \\
165 &   manga-8315-3703 &    35.5 &       10.3 &     5.0 &        1.7 &     146.6 &          5.6 &   0.139 &     11.1 \\
166 &   manga-8330-6104 &    40.1 &        9.1 &     4.7 &        1.2 &     139.8 &          6.1 &   0.139 &     10.8 \\
167 &   manga-8726-3701 &    59.0 &       10.8 &     5.1 &        1.1 &     170.9 &          7.7 &   0.140 &     11.2 \\
168 &   manga-8725-3704 &    41.8 &        7.5 &     4.3 &        1.0 &      71.4 &          5.9 &   0.141 &     10.5 \\
169 &   manga-8464-3701 &    23.7 &        5.4 &     4.3 &        1.4 &      37.9 &          5.1 &   0.142 &     10.4 \\
170 &   manga-8567-1901 &    30.1 &        6.3 &     1.8 &        0.5 &      49.3 &          4.4 &   0.142 &     10.0 \\
171 &   manga-8449-1901 &    66.2 &       11.3 &     2.1 &        0.5 &     165.7 &          8.2 &   0.142 &     10.7 \\
172 &   manga-8330-3703 &    52.8 &        7.5 &     4.6 &        0.8 &      69.6 &          6.1 &   0.143 &     10.8 \\
173 &   manga-8312-9102 &    58.3 &       12.5 &     2.6 &        0.5 &     151.1 &          4.2 &   0.143 &     10.8 \\
174 &   manga-8603-1902 &    68.9 &       10.7 &     1.9 &        0.3 &     134.2 &          6.3 &   0.143 &     10.7 \\
175 &   manga-8452-3702 &    57.5 &       14.3 &     5.6 &        1.8 &     215.5 &         14.9 &   0.144 &     11.1 \\
176 &   manga-8715-3704 &    44.0 &       14.5 &     3.7 &        1.2 &     235.7 &          5.7 &   0.144 &     11.1 \\
177 &   manga-8325-6104 &    80.5 &       10.7 &     5.2 &        1.0 &     161.8 &          9.6 &   0.145 &     11.1 \\
178 &   manga-8486-1902 &    99.5 &        8.6 &     1.2 &        0.1 &      64.5 &          4.6 &   0.145 &     10.0 \\
179 &   manga-8319-1901 &    38.5 &        9.2 &     2.1 &        0.5 &     112.8 &          5.2 &   0.145 &     10.4 \\
180 &   manga-8263-9102 &    52.4 &        9.8 &     5.7 &        1.4 &     155.3 &          7.3 &   0.146 &     11.0 \\
181 &   manga-8326-6103 &    84.7 &        7.8 &     4.4 &        0.4 &      74.7 &          4.0 &   0.146 &     10.6 \\
182 &   manga-8714-3702 &    75.0 &       11.8 &     4.9 &        1.1 &     191.9 &         10.4 &   0.148 &     10.9 \\
183 &   manga-8548-3701 &    31.7 &        9.1 &     5.3 &        1.7 &     134.8 &          5.7 &   0.149 &     11.0 \\
184 &   manga-8138-3704 &    67.4 &        6.4 &     1.9 &        0.2 &      41.9 &          3.6 &   0.150 &      9.9 \\
185 &  manga-10001-1902 &    66.4 &       10.4 &     2.2 &        0.5 &      75.2 &         12.1 &   0.151 &     10.2 \\
186 &   manga-8257-3701 &    55.1 &       10.4 &     3.2 &        0.8 &     154.7 &          7.8 &   0.151 &     11.0 \\
187 &   manga-8613-6104 &    45.2 &       10.7 &     7.8 &        2.2 &     166.0 &          9.5 &   0.151 &     11.3 \\
188 &   manga-7962-6103 &    76.6 &        9.8 &     6.5 &        0.9 &     139.1 &          6.2 &   0.153 &     11.2 \\
189 &   manga-8261-3704 &    38.8 &       11.7 &     7.4 &        2.8 &     149.9 &         11.8 &   0.153 &     11.2 \\
190 &  manga-8146-12705 &    76.5 &        7.3 &     5.5 &        0.7 &      70.1 &          4.5 &   0.154 &     10.9 \\
191 &   manga-8143-1901 &    45.2 &        8.2 &     2.6 &        0.6 &      51.0 &          8.9 &   0.155 &     10.1 \\
192 &   manga-8243-6104 &    53.4 &       10.9 &     4.2 &        1.1 &     188.3 &          8.2 &   0.155 &     10.9 \\
193 &   manga-8483-3701 &    85.4 &        7.5 &     2.0 &        0.2 &      43.7 &          3.2 &   0.155 &     10.5 \\
194 &   manga-8464-1902 &    74.7 &        8.4 &     3.1 &        0.5 &      54.0 &          9.1 &   0.155 &     10.7 \\
195 &   manga-8462-3702 &    49.9 &       11.3 &     8.2 &        2.3 &     211.0 &          9.3 &   0.156 &     11.3 \\
196 &   manga-8341-3701 &    95.6 &        9.0 &     2.7 &        0.3 &      73.2 &          4.1 &   0.157 &     10.6 \\
197 &   manga-8447-6102 &    76.8 &       11.2 &     7.0 &        1.1 &     185.0 &          5.9 &   0.157 &     11.3 \\
198 &   manga-8137-3703 &    52.7 &        8.8 &     2.1 &        0.4 &      86.9 &          5.1 &   0.157 &     10.6 \\
199 &   manga-7975-3702 &    53.6 &       11.7 &     6.0 &        1.6 &     215.5 &          8.9 &   0.157 &     11.3 \\
200 &   manga-8726-9101 &    60.0 &        9.6 &     5.3 &        1.1 &     135.3 &          6.9 &   0.158 &     11.1 \\
201 &   manga-8333-3704 &    46.6 &       12.4 &     7.0 &        2.3 &     242.6 &          9.2 &   0.159 &     11.3 \\
202 &   manga-8553-6103 &    81.1 &        7.8 &     3.9 &        0.4 &      72.1 &          4.5 &   0.159 &     10.7 \\
203 &   manga-8461-3701 &    74.4 &       14.1 &     6.9 &        1.6 &     133.6 &         15.2 &   0.159 &     10.9 \\
204 &   manga-8718-3702 &    57.0 &        9.8 &     3.2 &        0.6 &     112.3 &          4.8 &   0.160 &     10.9 \\
205 &   manga-8077-3702 &    43.2 &        7.0 &     1.9 &        0.4 &      62.6 &          5.7 &   0.160 &     10.2 \\
206 &  manga-8550-12701 &    53.1 &        9.2 &     4.7 &        1.1 &     134.1 &          6.9 &   0.160 &     11.0 \\
207 &   manga-8613-6103 &    43.7 &       10.1 &     2.7 &        0.7 &     143.8 &          5.2 &   0.160 &     10.6 \\
208 &   manga-8078-1902 &    61.6 &       11.4 &     1.2 &        0.2 &     108.4 &          4.2 &   0.160 &     10.4 \\
209 &   manga-8252-3702 &    80.0 &        7.0 &     6.5 &        0.8 &      61.5 &          5.6 &   0.161 &     10.8 \\
210 &   manga-8143-6104 &    56.0 &       10.3 &     4.7 &        1.2 &     167.8 &          8.1 &   0.161 &     10.9 \\
211 &   manga-8253-3704 &    28.8 &       11.2 &     1.9 &        0.7 &     146.6 &          4.7 &   0.161 &     10.5 \\
212 &   manga-8077-3704 &    43.0 &        9.9 &     2.7 &        0.8 &      68.5 &         11.1 &   0.161 &     10.3 \\
213 &   manga-8325-9102 &    48.6 &        6.6 &     6.0 &        0.9 &      64.3 &          4.0 &   0.162 &     10.9 \\
214 &   manga-8948-3702 &    55.9 &        8.3 &     1.7 &        0.3 &      70.1 &          3.6 &   0.163 &     10.4 \\
215 &   manga-8714-3704 &    48.4 &       11.5 &     5.3 &        1.9 &     151.1 &         12.6 &   0.163 &     11.0 \\
216 &   manga-8456-6103 &    38.5 &       10.1 &     7.5 &        2.2 &     158.3 &          6.5 &   0.163 &     11.3 \\
217 &  manga-9049-12705 &    87.4 &        9.6 &     3.9 &        0.5 &     110.3 &          4.8 &   0.163 &     11.0 \\
218 &   manga-8145-3702 &    29.5 &       12.3 &     0.6 &        0.2 &      47.9 &          1.8 &   0.163 &      9.6 \\
219 &   manga-8137-6103 &    64.9 &       12.5 &     5.5 &        1.6 &     155.3 &         13.9 &   0.164 &     11.0 \\
220 &   manga-8078-1901 &    30.6 &        8.7 &     2.1 &        0.9 &     105.7 &          8.3 &   0.164 &     10.3 \\
221 &   manga-8547-1902 &    42.5 &       10.5 &     5.8 &        1.6 &     162.5 &          6.0 &   0.164 &     11.1 \\
222 &   manga-8326-3701 &   106.0 &        9.0 &     2.0 &        0.2 &      78.8 &          5.6 &   0.164 &     10.4 \\
223 &   manga-8718-6101 &    70.5 &        8.1 &     6.4 &        0.9 &      82.1 &          6.1 &   0.165 &     10.9 \\
224 &   manga-8259-3704 &    52.9 &        8.8 &     6.9 &        1.6 &      75.8 &          9.1 &   0.167 &     10.9 \\
225 &   manga-8330-6101 &    77.7 &        8.4 &     3.8 &        0.4 &      91.6 &          4.4 &   0.167 &     10.7 \\
226 &   manga-8485-6104 &    53.7 &        9.8 &     4.2 &        0.8 &     139.3 &          5.4 &   0.168 &     10.9 \\
227 &   manga-8454-3703 &    55.6 &        9.9 &     6.1 &        1.3 &     156.8 &          7.5 &   0.169 &     10.9 \\
228 &   manga-8550-1901 &    78.0 &        8.1 &     1.9 &        0.2 &      68.7 &          6.4 &   0.170 &     10.1 \\
229 &   manga-8548-1902 &    48.4 &        6.7 &     1.2 &        0.2 &      36.4 &          2.8 &   0.171 &      9.5 \\
230 &   manga-8456-3704 &    23.9 &        9.1 &     1.4 &        0.4 &      57.0 &          2.7 &   0.171 &     10.0 \\
231 &   manga-8319-1902 &    76.0 &        7.9 &     1.7 &        0.2 &      58.6 &          5.1 &   0.171 &     10.1 \\
232 &   manga-8604-1901 &    58.0 &       10.7 &     2.3 &        0.6 &     137.7 &         10.8 &   0.172 &     10.5 \\
233 &   manga-8315-1901 &    53.3 &        6.9 &     2.0 &        0.3 &      53.9 &          3.6 &   0.172 &     10.0 \\
234 &   manga-8319-6103 &    51.9 &        7.3 &     5.6 &        1.0 &      84.5 &          5.8 &   0.174 &     10.9 \\
235 &  manga-8553-12704 &    52.8 &        7.5 &     5.1 &        0.8 &      67.5 &          3.2 &   0.176 &     10.9 \\
236 &   manga-8949-6102 &    58.8 &        6.6 &     1.8 &        0.2 &      42.0 &          3.5 &   0.176 &     10.2 \\
237 &   manga-8931-3704 &    54.9 &        6.6 &     1.3 &        0.2 &      39.9 &          2.9 &   0.177 &      9.3 \\
238 &  manga-10001-6103 &    92.1 &        7.6 &     3.3 &        0.3 &      58.1 &          4.9 &   0.177 &     10.7 \\
239 &  manga-8140-12701 &    86.8 &        6.8 &     3.0 &        0.3 &      46.0 &          3.4 &   0.177 &     10.5 \\
240 &   manga-8261-9101 &   103.6 &       24.3 &    16.7 &        4.3 &      83.0 &         26.5 &   0.178 &     11.0 \\
241 &   manga-8133-3703 &    74.0 &       10.6 &     4.6 &        0.8 &     148.7 &          9.4 &   0.179 &     10.8 \\
242 &   manga-8715-3703 &    45.4 &        9.0 &     3.6 &        0.8 &     114.3 &          5.5 &   0.179 &     10.7 \\
243 &   manga-7991-6101 &    90.7 &        9.2 &     2.2 &        0.2 &      79.4 &          4.9 &   0.179 &     10.6 \\
244 &   manga-8552-6104 &    44.4 &       10.6 &     5.1 &        1.5 &     177.4 &          6.9 &   0.179 &     11.1 \\
245 &   manga-8440-3701 &    70.2 &       10.5 &     2.0 &        0.3 &     124.6 &          5.2 &   0.180 &     10.3 \\
246 &   manga-9049-3704 &    59.6 &        8.5 &     1.8 &        0.3 &      91.8 &          6.9 &   0.181 &      9.6 \\
247 &   manga-9049-1901 &    67.5 &        9.6 &     7.0 &        1.2 &      91.5 &          6.1 &   0.181 &     11.2 \\
248 &   manga-8611-3701 &   107.7 &        7.9 &     2.8 &        0.3 &      63.5 &          5.4 &   0.182 &     10.6 \\
249 &   manga-8721-6103 &   113.7 &        9.3 &     3.5 &        0.3 &      74.6 &          3.6 &   0.182 &     10.9 \\
250 &   manga-8718-1901 &    51.5 &       10.6 &     2.7 &        0.8 &     139.7 &         10.5 &   0.184 &     10.8 \\
251 &  manga-8604-12705 &    95.4 &        7.4 &     4.0 &        0.4 &      55.5 &          6.0 &   0.184 &     10.7 \\
252 &   manga-9042-3704 &    48.4 &       14.0 &     4.1 &        1.7 &     205.6 &         15.3 &   0.185 &     10.9 \\
253 &   manga-8724-6102 &    63.2 &        5.9 &     3.1 &        0.3 &      34.2 &          2.5 &   0.185 &     10.3 \\
254 &  manga-7495-12701 &    83.9 &        8.4 &     4.4 &        0.4 &      72.3 &          3.3 &   0.185 &     10.8 \\
255 &   manga-8603-9101 &    90.9 &        7.4 &     2.5 &        0.2 &      46.8 &          3.5 &   0.186 &     10.5 \\
256 &   manga-8950-1902 &    90.5 &       10.0 &     1.3 &        0.2 &      92.5 &          4.9 &   0.186 &     10.2 \\
257 &   manga-8254-3702 &    43.9 &        8.6 &     2.7 &        0.6 &     101.3 &          4.7 &   0.187 &     10.5 \\
258 &  manga-8602-12705 &    98.2 &        7.5 &     3.7 &        0.3 &      55.9 &          3.8 &   0.188 &     10.8 \\
259 &   manga-8252-3704 &    70.5 &        8.1 &     3.1 &        0.4 &      82.3 &          4.8 &   0.188 &     10.6 \\
260 &   manga-8450-6104 &    78.5 &       10.2 &     2.9 &        0.4 &      99.9 &          3.9 &   0.189 &     11.0 \\
261 &   manga-8319-3702 &    57.0 &        7.6 &     2.9 &        0.4 &      75.9 &          3.9 &   0.190 &     10.5 \\
262 &   manga-8329-1901 &    84.3 &        9.4 &     1.7 &        0.2 &      96.9 &          6.0 &   0.191 &     10.4 \\
263 &   manga-8978-1901 &    58.5 &        8.5 &     2.5 &        0.5 &      75.9 &          8.5 &   0.192 &     10.5 \\
264 &   manga-8592-1901 &    65.3 &        8.8 &     2.3 &        0.4 &      79.4 &          9.0 &   0.192 &     10.1 \\
265 &   manga-8253-9101 &    52.2 &       10.1 &     3.6 &        0.9 &     147.3 &          6.3 &   0.193 &     10.9 \\
266 &   manga-8604-6104 &    45.0 &        8.7 &     2.9 &        0.8 &     116.0 &          7.2 &   0.194 &     10.8 \\
267 &   manga-8335-6101 &    42.2 &       11.6 &     2.3 &        0.8 &      72.4 &         12.8 &   0.194 &     10.1 \\
268 &   manga-8147-3704 &    62.1 &        9.2 &     1.9 &        0.3 &      86.3 &          4.0 &   0.195 &     10.6 \\
269 &   manga-8454-3701 &    56.5 &       12.3 &     5.8 &        1.7 &     117.5 &         13.6 &   0.195 &     11.0 \\
270 &   manga-8455-1901 &    23.5 &        6.0 &     3.0 &        1.1 &      59.9 &          5.3 &   0.195 &      9.8 \\
271 &   manga-8135-1901 &    51.1 &        8.1 &     2.9 &        0.7 &      59.0 &          9.0 &   0.196 &      9.8 \\
272 &   manga-8464-3702 &    33.8 &        8.1 &     1.8 &        0.5 &      82.5 &          3.8 &   0.196 &     10.5 \\
273 &   manga-8259-3702 &    40.1 &        6.4 &     2.8 &        0.6 &      50.0 &          4.3 &   0.196 &     10.5 \\
274 &   manga-8243-3704 &    53.0 &       11.5 &     1.9 &        0.4 &     131.9 &          4.4 &   0.196 &     10.4 \\
275 &   manga-8459-3703 &    79.3 &        9.7 &     3.6 &        0.4 &      65.0 &          2.7 &   0.197 &     11.0 \\
276 &  manga-9041-12704 &    76.6 &        6.6 &     8.5 &        1.1 &      70.4 &          3.8 &   0.198 &     11.0 \\
277 &   manga-8155-6103 &    81.6 &        9.6 &     6.7 &        0.8 &      99.7 &          4.7 &   0.199 &     11.3 \\
278 &   manga-8724-3702 &    69.2 &       10.6 &     2.3 &        0.4 &     133.3 &          5.4 &   0.199 &     10.7 \\
279 &   manga-8979-3704 &    68.3 &        9.2 &     4.2 &        0.5 &      74.4 &          3.4 &   0.199 &     10.7 \\
280 &   manga-8588-6101 &    38.8 &       10.4 &     2.3 &        0.5 &      88.2 &          3.1 &   0.199 &     11.0 \\
281 &   manga-8455-3704 &    62.6 &        7.7 &     3.1 &        0.5 &      38.8 &          7.7 &   0.200 &      9.9 \\
282 &   manga-8482-3701 &    54.7 &       11.4 &     2.4 &        0.5 &     127.8 &          4.2 &   0.201 &     10.7 \\
283 &   manga-8258-3703 &    63.6 &       11.2 &     3.9 &        0.7 &     140.6 &          5.1 &   0.204 &     11.0 \\
284 &   manga-8243-3703 &    77.0 &        6.4 &     2.7 &        0.3 &      43.3 &          3.3 &   0.204 &     10.3 \\
285 &   manga-8312-1902 &    62.0 &        9.8 &     2.0 &        0.4 &     110.0 &          5.6 &   0.206 &     10.4 \\
286 &  manga-8728-12705 &    43.4 &        5.8 &     6.9 &        1.1 &      34.6 &          3.5 &   0.206 &     10.8 \\
287 &   manga-8597-1902 &    80.8 &        7.8 &     2.6 &        0.3 &      58.1 &          5.1 &   0.207 &     10.4 \\
288 &  manga-8934-12705 &    45.5 &        6.5 &     2.5 &        0.4 &      51.3 &          3.2 &   0.209 &     10.3 \\
289 &  manga-8947-12704 &    69.3 &        6.3 &     4.3 &        0.6 &      38.4 &          5.4 &   0.210 &     10.8 \\
290 &   manga-8600-3702 &    80.3 &        8.7 &     2.5 &        0.3 &      86.1 &          5.4 &   0.210 &     10.4 \\
291 &  manga-8946-12702 &    91.9 &        9.8 &     6.7 &        1.0 &      60.4 &         10.7 &   0.211 &     11.0 \\
292 &   manga-8243-1902 &    63.9 &        9.4 &     2.9 &        0.6 &     130.5 &          8.4 &   0.211 &     10.2 \\
\bottomrule
\end{longtable}
\twocolumngrid

The characterization of the environment for the MaNGA sample is composed of two main parameters, namely the local number density to the fifth nearest neighbor, $\eta$, and the tidal strength affecting the primary galaxy, $Q$, as described in Argudo-Fernandez et al. (2013, 2014, 2015) where it was established that the structure of a galaxy may be affected by external influences when the corresponding tidal force amounts to $>$1\% of the internal binding force, corresponding to a critical tidal strength of $Q_{crit}=-2$ and a critical local number density of $\eta_{crit}=2.7$. The mean tidal strength and local number density values for our sub sample of 292 low rotating ellipticals are $Q_{mean}=-2.34$ and $\eta_{mean}=0.59$, well below the critical interaction values, with maximum values reaching $Q_{max}=0.63$ and $\eta_{max}=2.12$. Thus, our mean results should be robust to tidal effects, while a certain spread coming from high $Q$ values might be expected. Further, these indexes use only projected distances, and are therefore upper limits on the true effects. We can therefore be confident that being below the critical thresholds on the average shows our results will not be driven by tidal galaxy interactions.

These isolation criteria exclude systems where the Newtonian tides (assuming the line of sight separation to be zero) are above 1\% of the internal binding forces at the outskirts, this also implies that for equal masses for the target and the closest neighbor (a safe upper limit, as the sample construction criteria imply a factor of 4 reduction in luminosity between the target galaxy and its closest neighbor), the external acceleration will be about $100^{1/3} = 4.64$ times smaller than the internal one, assuming both to be in the deep MOND regime. This last condition is a reasonable approximation, as the internal accelerations at the outskirts do fall below the $a_0$ threshold. Thus, the sample construction ensures that any contribution from an external field effect will quite probably
fall below the other internal uncertainties of our experiment, e.g. those given by the confidence intervals of our velocity dispersion profile observations and fits.

A random selection of nine fitted velocity dispersion profiles is presented in figure 1. The adequacy of the fits is evident despite the variety of central concentration in observed kinematic profile morphologies.

\begin{figure}
\plotone{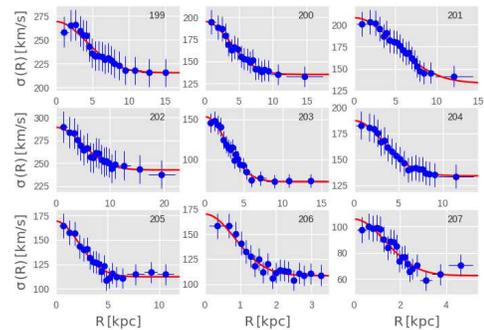}
\caption{Projected velocity dispersion profiles for the 9 random elliptical galaxies of our sub sample, as a function of radial distance in the system, with vertical error bars showing the averaged empirical errors.
The top right number of each profile gives the identification number for reference in table 1.
The solid curve gives the best fit to the universal profile proposed. The complete figure set of the 292 elliptical galaxies studied can be found in the online published version of this article.}
\end{figure}

An initial examination of our first sample showed significant scatter in the $R_{\sigma}$ to $\sigma_{\infty}$ relation, which inspired us to analyze the squared sum of relative errors:

\begin{equation}
\delta_{rel}= \frac{\delta \sigma_0}{\sigma_0} + \frac{\delta R_{\sigma}}{R_{\sigma}} +  \frac{\delta \sigma_{\infty}}{\sigma_{\infty}} = \sum_{i=1}^{3} \frac{\delta x_i}{x_i}
\end{equation}

\noindent where $x_i$ represents each fitted parameter and $\delta x_i$ the respective uncertainty. Next, we select a subsequent sub-sample, chosen as the first quintile of the distribution of summed relative errors, in order to study and compare the behavior of our larger sample. The second sub-sample comprises the 60 systems with the lowest sum of relative errors (table 2). Figure 2 shows the fitted projected velocity dispersion profiles of a random selection of 9 galaxies from our second sub-sample of lowest relative error objects. Noteworthy is the evident reduction in uncertainties and overall clearer profile morphology. 
In the following section we compare this second-sub sample with the larger collection of galaxies and perform various comparisons between the fitted parameters and the stellar mass of each object as estimated from the MaNGA survey, in order to test the consistency of our samples.

We use the measured redshift from the MaNGA source catalog, and adopt stellar masses estimated from Sanchez et al. (2017a).
A detailed description of the selection parameters can be found in Bundy et al. (2015), Law et al. (2016) and Yan et al. (2016b), as well as a description of sample properties in Wake et al. (2017).

\begin{figure}
\plotone{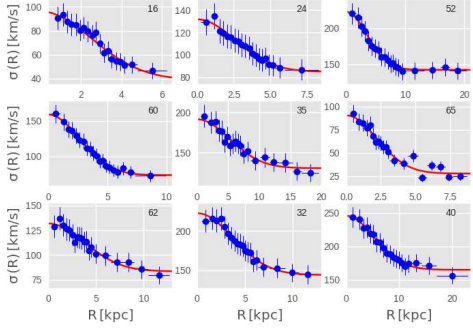}
\caption{A random selection of 9 fitted velocity dispersion profiles from our lowest relative error sub sample. Blue dots represent the averaged velocity dispersion observations with horizontal error bars delimiting each radial bin, and vertical error bars representing the propagated empirical errors. The red curves are the fit to eq. (6).}
\end{figure}

~\\ ~\\ ~\\

\onecolumngrid

\begin{longtable}{lccrrccc}
\caption{Fitted parameters with respective errors, as well as the sum of the squared relative errors for the lowest relative error sub sample. Galaxy ID corresponds to ID and MaNGA names from table 1.}\\
\toprule
ID &  $\sigma_0$ &  $\delta\sigma_0$ &  $R_{\sigma}$ &  $\delta R_{\sigma}$ &  $\sigma_{\infty}$ &  $\delta\sigma_{\infty}$ &  $\sum_{i=1}^3 (\frac{\delta\/x_i}{x_i})^2$ \\
{} & $[km/s]$ & $[km/s]$ & $[kpc]$ & $[kpc]$ & $[km/s]$ & $[km/s]$ & {} \\
\midrule
\endhead
77  &   113.7 &        9.3 &     3.5 &        0.3 &      74.6 &          3.6 &   0.016 \\
36  &    98.2 &        7.5 &     3.7 &        0.3 &      55.9 &          3.8 &   0.017 \\
24  &    86.8 &        6.8 &     3.0 &        0.3 &      46.0 &          3.4 &   0.019 \\
26  &    90.9 &        7.4 &     2.5 &        0.2 &      46.8 &          3.5 &   0.020 \\
81  &   110.5 &        8.8 &     2.1 &        0.2 &      75.2 &          4.9 &   0.020 \\
73  &    95.6 &        9.0 &     2.7 &        0.3 &      73.2 &          4.1 &   0.021 \\
78  &    84.7 &        7.8 &     4.4 &        0.4 &      74.7 &          4.0 &   0.021 \\
21  &    77.0 &        6.4 &     2.7 &        0.3 &      43.3 &          3.3 &   0.021 \\
69  &    83.9 &        8.4 &     4.4 &        0.4 &      72.3 &          3.3 &   0.021 \\
51  &    99.5 &        8.6 &     1.2 &        0.1 &      64.5 &          4.6 &   0.022 \\
48  &   114.1 &        8.0 &     4.8 &        0.4 &      63.9 &          5.9 &   0.022 \\
17  &    96.9 &        7.2 &     3.2 &        0.3 &      41.4 &          3.8 &   0.022 \\
64  &    88.7 &        8.9 &     3.5 &        0.3 &      70.2 &          3.5 &   0.022 \\
22  &    85.4 &        7.5 &     2.0 &        0.2 &      43.7 &          3.2 &   0.022 \\
25  &    67.7 &        6.1 &     5.5 &        0.6 &      46.3 &          2.7 &   0.022 \\
47  &   107.7 &        7.9 &     2.8 &        0.3 &      63.5 &          5.4 &   0.022 \\
88  &   106.0 &        9.0 &     2.0 &        0.2 &      78.8 &          5.6 &   0.023 \\
68  &    81.1 &        7.8 &     3.9 &        0.4 &      72.1 &          4.5 &   0.024 \\
5   &    63.2 &        5.9 &     3.1 &        0.3 &      34.2 &          2.5 &   0.024 \\
39  &    92.1 &        7.6 &     3.3 &        0.3 &      58.1 &          4.9 &   0.024 \\
1   &    62.7 &        5.3 &     3.1 &        0.3 &      27.9 &          2.4 &   0.024 \\
65  &    76.6 &        6.6 &     8.5 &        1.1 &      70.4 &          3.8 &   0.025 \\
62  &    76.5 &        7.3 &     5.5 &        0.7 &      70.1 &          4.5 &   0.027 \\
18  &    67.4 &        6.4 &     1.9 &        0.2 &      41.9 &          3.6 &   0.027 \\
90  &    90.7 &        9.2 &     2.2 &        0.2 &      79.4 &          4.9 &   0.027 \\
110 &    77.7 &        8.4 &     3.8 &        0.4 &      91.6 &          4.4 &   0.027 \\
128 &    87.4 &        9.6 &     3.9 &        0.5 &     110.3 &          4.8 &   0.028 \\
112 &    90.5 &       10.0 &     1.3 &        0.2 &      92.5 &          4.9 &   0.029 \\
52  &    79.3 &        9.7 &     3.6 &        0.4 &      65.0 &          2.7 &   0.029 \\
35  &    95.4 &        7.4 &     4.0 &        0.4 &      55.5 &          6.0 &   0.029 \\
45  &    80.0 &        7.0 &     6.5 &        0.8 &      61.5 &          5.6 &   0.030 \\
74  &    85.0 &        8.1 &     3.7 &        0.5 &      73.8 &          5.9 &   0.030 \\
2   &    66.6 &        5.8 &     3.0 &        0.3 &      30.3 &          3.5 &   0.030 \\
117 &    81.6 &        9.6 &     6.7 &        0.8 &      99.7 &          4.7 &   0.031 \\
38  &    80.8 &        7.8 &     2.6 &        0.3 &      58.1 &          5.1 &   0.031 \\
72  &    70.8 &        8.7 &     4.7 &        0.5 &      73.0 &          3.5 &   0.031 \\
3   &    61.2 &        6.1 &     2.1 &        0.2 &      31.2 &          3.2 &   0.032 \\
228 &    92.4 &       11.7 &     5.5 &        0.7 &     166.0 &          4.8 &   0.032 \\
101 &    80.3 &        8.7 &     2.5 &        0.3 &      86.1 &          5.4 &   0.033 \\
92  &    70.5 &        8.1 &     3.1 &        0.4 &      82.3 &          4.8 &   0.033 \\
60  &    79.3 &        7.5 &     5.3 &        0.7 &      69.9 &          6.3 &   0.034 \\
175 &    80.5 &       10.2 &     5.3 &        0.7 &     143.0 &          5.7 &   0.035 \\
19  &    58.8 &        6.6 &     1.8 &        0.2 &      42.0 &          3.5 &   0.035 \\
40  &    76.0 &        7.9 &     1.7 &        0.2 &      58.6 &          5.1 &   0.035 \\
114 &    84.3 &        9.4 &     1.7 &        0.2 &      96.9 &          6.0 &   0.035 \\
58  &    78.0 &        8.1 &     1.9 &        0.2 &      68.7 &          6.4 &   0.036 \\
16  &    54.9 &        6.6 &     1.3 &        0.2 &      39.9 &          2.9 &   0.036 \\
119 &    78.5 &       10.2 &     2.9 &        0.4 &      99.9 &          3.9 &   0.037 \\
91  &    70.5 &        8.1 &     6.4 &        0.9 &      82.1 &          6.1 &   0.037 \\
76  &    68.3 &        9.2 &     4.2 &        0.5 &      74.4 &          3.4 &   0.037 \\
142 &    74.5 &        8.9 &     5.4 &        0.8 &     120.2 &          6.4 &   0.038 \\
144 &    95.7 &       14.3 &     1.4 &        0.2 &     124.3 &          3.6 &   0.039 \\
166 &    76.6 &        9.8 &     6.5 &        0.9 &     139.1 &          6.2 &   0.039 \\
83  &    61.7 &        7.7 &     4.5 &        0.6 &      75.8 &          5.4 &   0.040 \\
124 &    78.0 &        9.5 &     2.8 &        0.4 &     107.5 &          5.4 &   0.040 \\
174 &    80.1 &       10.6 &     4.9 &        0.7 &     142.4 &          5.1 &   0.040 \\
20  &    58.0 &        6.2 &     5.2 &        0.7 &      42.0 &          4.7 &   0.040 \\
84  &    57.0 &        7.6 &     2.9 &        0.4 &      75.9 &          3.9 &   0.041 \\
4   &    77.2 &        6.7 &     3.9 &        0.4 &      33.6 &          5.1 &   0.041 \\
33  &    53.3 &        6.9 &     2.0 &        0.3 &      53.9 &          3.6 &   0.042 \\
\bottomrule
\end{longtable}
\twocolumngrid

~\\
\section{Comparisons with MONDian expectations}

In his original paper, Milgrom (1983) indicated that MOND suggests a mass-velocity dispersion relation for elliptical galaxies of the form $M\propto \sigma^4$. If there were no systematic variation of the mass-to-light ratio (ML), this would become the observed Faber-Jackson (FJ) relation (Faber \& Jackson 1976). Even though the main interest of this study focuses on eq. (4) in which we have sidestepped the need for a mass calculation, we nevertheless seek to test consistency with previous studies using the stellar mass estimates provided by the MaNGA team.

Figure (3) shows the central velocity dispersion, $\sigma (0)=\sigma_0 + \sigma_{\infty}$, as a function of stellar mass and $\delta_{Rel}$, the total sum of the relative errors of the fitted parameters. The black dashed line represents precisely the $M\propto \sigma^4$ FJ relation (Sanders 2010). There is a good agreement between the FJ proportionality and our sample, with some expected scatter. Interestingly, systems on the top right of the relation present $\delta_{rel}$ values an order of magnitude higher than galaxies on the bottom left of the spectrum. As previously mentioned, this is due to an intrinsic redshift bias in the survey, in which the galaxies with greater absolute magnitude are also the ones with higher redshift, leading to systematically greater observational errors. In figure (1) of Sanchez et al. (2017b), this bias can be observed for systems in the redshift range $z>0.06$, while systems below this redshift value seem to consolidate a more complete sample.

\begin{figure}
\plotone{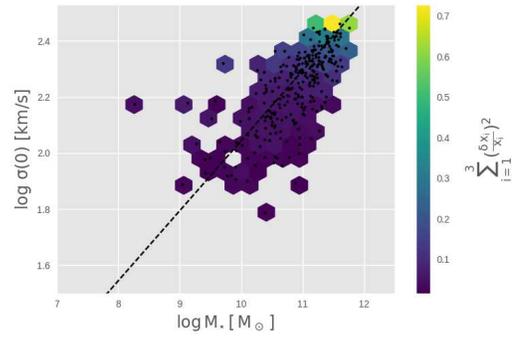}
\caption{A hexbin plot of the central velocity dispersion as a function of stellar mass. The color palette represents the third plotted variable, the total sum of the relative errors of the fitted parameters. The black dots show the actual sample distribution. The black dashed line represents the $M\propto \sigma^4$ Faber-Jackson proportionality. A good fit is evident, as is the tendency for low mass and low velocity dispersion objects to have the least relative errors.}
\end{figure}

 A similar comparison is made in figure (4), this time relating the asymptotic velocity dispersion $\sigma_{\infty}$ to stellar mass. Again we find that higher mass and asymptotic velocity dispersion objects display the largest relative errors, with lower mass and $\sigma_{\infty}$ systems the smallest relative errors. It is evident that $\sigma_{\infty}$ follows the FJ scaling of figure (3) for high masses, though with a scaled down amplitude. Indeed, a good fit to the expected MONDian Tully-Fisher expectations of the dotted line (eq. 2) is obvious above $\sigma_{\infty}=100 km/s$. Below this value, a much more scattered situation appears, possibly due to uncertainties in the stellar mass determinations.

\begin{figure}
\plotone{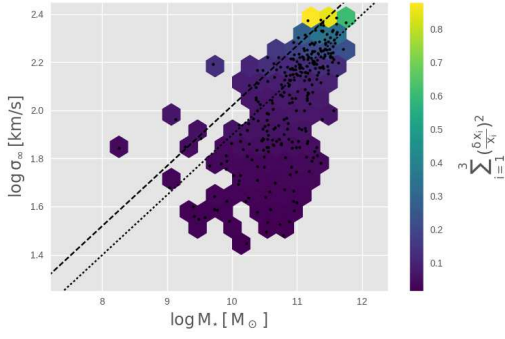}
\caption{A hexbin plot of the asymptotic velocity dispersion as a function of stellar mass. The color palette represents the third plotted variable, the total sum of the relative errors of the fitted parameters. The black dots show the actual sample distribution. The black dashed line represents the $M\propto \sigma^4$ proportionality of the FJ relation from figure (3). The black dotted line gives the Tully-Fisher relation for pressure supported systems in a MONDian regime. A deviation from a Faber-Jackson like tendency can be observed for low mass and low velocity dispersion objects.}
\end{figure}

We now study the effects of central concentration and velocity dispersion profile morphology by defining the relative velocity dispersion of a system:

\begin{equation}
\sigma_{Rel} = \frac{\sigma_0 + \sigma_{\infty}}{\sigma_{\infty}}
\end{equation}

\noindent and comparing it to the stellar mass of each galaxy (figure 5). A higher $\sigma_{Rel}$ value denotes a larger central velocity dispersion, $\sigma (0)$ with respect to the outer asymptotic velocity dispersion $\sigma_{\infty}$, i.e. a steeper drop in the amplitude of the velocity dispersion profile. It is evident from figure (5) that the more massive galaxies show the largest relative errors, while galaxies with a more pronounced fall in the velocity dispersion profile correspond to the lower masses and lower relative errors. 
Galaxies in the high end tail of the FJ proportionality in figure (4) best fit the $M\propto \sigma^4$ proportionality and also lie in the bottom right end of figure (5), in all likelihood due to the fact that we are observing the very central regions of these higher redshift objects, leading to higher errors and asymptotic velocity dispersions. Noteworthy from figure (5) is that no clear relationship is apparent  between $\sigma_{Rel}$ and $M_{\star}$, implying that no single univariate profile function of stellar mass can be expressed, needing at least two independent parameters, although mass dependent errors and varying radial coverage are a caveat for the larger systems. 

In going from local volumetric velocity dispersion values to the observed projected velocity dispersion profiles, different projection effects arise, which depend on the degree of central concentration of the volumetric velocity dispersion profile, and the real space density profiles of the tracers being
used e.g. Hernandez \& Jimenez (2012), Jimenez et al. (2013), Tortora (2014). Still, taking the simple identification of $R_{\sigma}=R_M$ introduced in Durazo et al. (2017), and building on the main result of our previous work, figure (6) relates the asymptotic velocity dispersion, $\sigma_{\infty}$, with the flattening radii, $R_{\sigma}$, both as a function of their respective summed relative errors. Black dots show the actual sample distribution and the solid black line represents the MONDian prediction of equation (4). We can see from this result that the systems which best match the expected scaling of eq. (4) are the low $\delta_{Re;}$ ones, which as we saw earlier also correspond to the lower mass, relative velocity dispersion and asymptotic velocity dispersion galaxies. Objects that scatter above the expected MONDian relation tend to present the highest $\delta_{Rel}$ and be the most massive galaxies.

\begin{figure}
\plotone{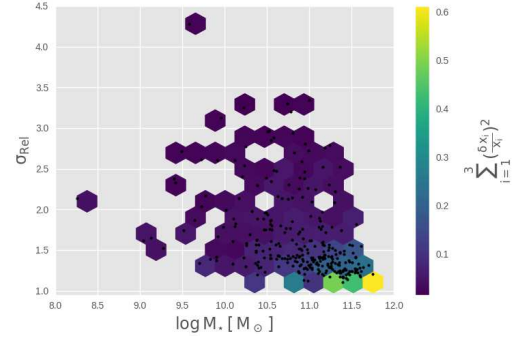}
\caption{A hexbin plot of the relative velocity dispersion as a function of stellar mass. The color palette represents the third plotted variable, the total sum of the relative errors of the fitted parameters. The black dots show the actual sample distribution. More massive galaxies show the highest relative errors, while those with steeper and more differentiated velocity dispersion profiles show the least sum of relative errors. No clear trend can be observed between $\sigma_{Rel}$ and $M_{\star}$. This shows the need for at least 2 parameters in describing the kinematics of the studied systems.}
\end{figure}

\begin{figure}
\plotone{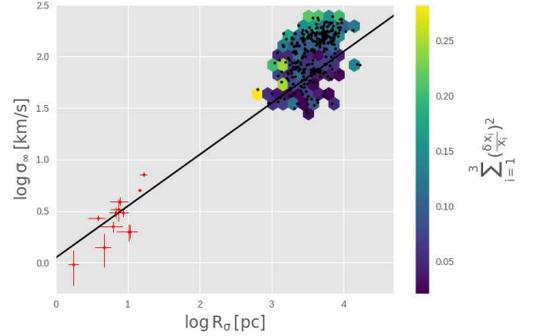}
\caption{$\sigma_{\infty}$ vs $R_{\sigma}$ values for our larger sample of 292 low rotating ellipticals. The color palette represents the third plotted variable, the total sum of the relative errors of the fitted parameters. The black dots show the actual sample distribution. Points at the lower left of the plot show the GC velocity profiles studied in Durazo et al. (2017). The solid line is not a fit to the data, but actually shows the MONDian expectations of equation (4) for the predicted scaling of $R_{M}=3 \sigma_{\infty}^{2}/a_{0}$, $R_{M}/pc = 0.81 (\sigma_{\infty}/kms^{-1})^{2}$}
\end{figure}

Finally in figure (7), we again display  $\sigma_{\infty}$ as a function of $R_{\sigma}$, this time for the first quintile of $\delta_{Rel}$, i.e. our second-sub sample containing the 60 objects with the lowest sum of squared relative errors. The consistency with the MONDian expectation of equation (4) is clear, especially considering that we are only plotting the systems within the first quintile of summed quadratic errors. Overall the match for both sub-samples is quite impressive given the sample size and tighter statistical treatment with respect to Durazo et al. (2017).
For this sub-sample, the ambient density parameters have values of $\eta_{mean}=0.63$ and $\eta_{max}=1.98$ while $Q$ values show a mean of $Q_{mean}=-1.97$, all well below the 1\% tidal effect threshold of $\eta_{crit}=2.7$ and $Q_{crit}=-2$, guaranteeing that the means of our results in figure (7) are not affected by tidal effects. However, the distribution of Q values reaches $Q_{max}=-0.43$, showing that some of the dispersion in the figure could well be due to systems with a certain degree of tidal interactions.

Figures 6 and 7 also include a sample of Galactic globular clusters with measured radial projected velocity dispersion profiles out to several half light radii, and well measured proper motions in the outer regions, from data produced by Scarpa et al. (2007a,b), Scarpa \& Falomo (2010), Scarpa et al. (2011), and Lane et al. (2009, 2010a,b, 2011). These are included as they provide a second set of systems several orders of magnitude away from the ellipticals we primarily study, and which are hence important to validate the comparison with a MOND-like behavior, not only in the amplitude of the $\sigma_{\infty}$ vs. $R_{\sigma}$ relation, but also in terms of the slope of this relation. Despite the evident agreement with the simple first order MONDian expectations, it must be noted that in the context of MOND as such, the external field effect would preclude the appearance of the observed flattening in the velocity dispersion profiles of these clusters, which hence renders their relevance subject to revision in terms of what the particular external field effect might be in any final covariant theory having a MONDian low velocity limit. The inclusion of Galactic globular clusters in these two figures is therefore tentative and at this point would imply placing oneself within the context of a fiducial MONDian theory without any external field effect, or one with a very limited such effect, e.g. Milgrom (2011).

As a final consistency check, we plot in figure (8) $\delta\/R_{\sigma}/R_{\sigma}$ values as a function of the field of view, i.e. the ratio of the aperture radius to effective radius, and indeed find an evident anti-correlation between the $R_{\sigma}$ relative errors (the parameter in figures 5, 6 and 7 with the largest uncertainties) and the extension to which each galaxy is observed, providing justification for our sub-sample selection from figure (7). Furthermore, deviations from MONDian expectations for the highest mass systems could be the result of a similar flattening in the velocity dispersion profiles before reaching a deep-MOND regime, as seen by Richtler et al. (2011) in circular velocity profiles of elliptical galaxies modeled by a Jaffe mass profile, pointing to the potential caveat of assuming that the enclosed baryonic mass has converged upon reaching a flattened velocity dispersion profile.

It seems that a distinctive gravitational physics applies, with a clear transition radius at $R_{M}$ where the ``Keplerian decline'' of the inner Newtonian $R<R_{M}$ region gives way to the
``Tully-Fisher'' $\sigma(R) \propto (G M a_{0})^{1/4}$ of the outer MONDian regime,  notwithstanding the large sample size. 

\begin{figure}
\plotone{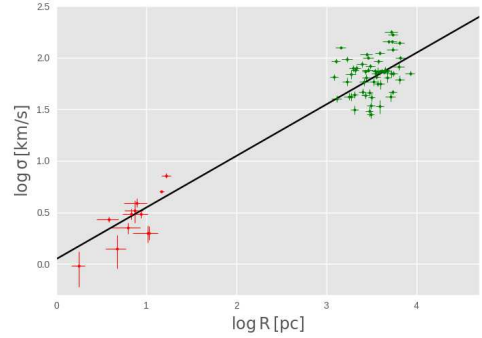}
\caption{$\sigma_{\infty}$ vs $R_{\sigma}$ values for our sub-sample of the first quintile of lowest summed relative errors. Points at the lower left of the plot show the GC velocity profiles studied in Durazo et al. (2017). The solid line is not a fit to the data, but actually shows the MONDian expectations of equation (4) for the predicted scaling of $R_{M}=3 \sigma_{\infty}^{2}/a_{0}$, $R_{M}/pc = 0.81 (\sigma_{\infty}/kms^{-1})^{2}$ through the identification of $R_{\sigma}=R_M$.}
\end{figure}

\begin{figure}
\plotone{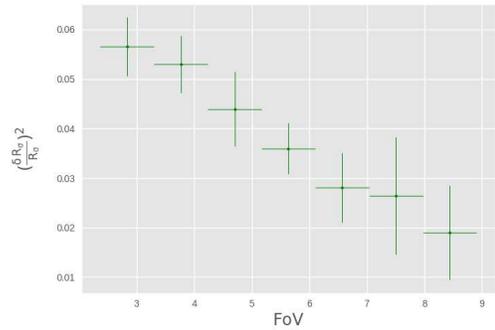}
\caption{$\delta R_{\sigma}/R_{\sigma}$ values as a function of the relative field of view for the larger 292 object sub sample. A clear anti-correlation between this parameter and the extent to which each galaxy is observed is evident, justifying our second sub sample selection.}
\end{figure}

Creating a similar large sample selection of isolated Galactic globular clusters at large Galactic radii, where the interpretation under MONDian theories becomes less contentious due to a negligible field effect, to solidify our original results is evidently a desirable development, which however is not an easy endeavor due to the need for independent stellar synthesis models or well measured proper motions of the most external regions of each cluster, to exclude Galactic tidal heating alternatives. Also, a sample of radially resolved kinematic observations for other pressure supported non relativistic astrophysical systems, such as dwarf spheroidals and ultra-faint dwarfs, would be appealing.\\

~\\
\section{Conclusions}\label{ccl}
We have once again shown that the universal velocity dispersion profile proposed in eq. (6) accurately models pressure supported elliptical galaxies showing an inner Keplerian decline and an outer flattening in their velocity dispersion values. 
We also find in general a good agreement with MONDian predictions for the whole sample, with an expected scatter for the brighter, more massive and higher redhisft galaxies, stemming back from the original MaNGA sample selection. Since higher $\sigma_{\infty}$ systems on average are larger galaxies which are more centrally observed, and hence observations do not reach the outer low acceleration regime, the clear deviations from modified gravity expectations shown in figure (6) can be explained by the smaller relative radii at which the velocity dispersion measurements are made for larger galaxies. 

Meanwhile, the low mass systems, presenting the lowest relative errors, show much better consistency with the expectations from MOND, in particular our least relative error sample shows  excellent compatibility with the modified gravity prediction. The above in spite of projection effects inevitably leading to a spread in parameters in going from the volumetric predictions of equation (4) to the projected quantities seen in figures 6 and 7.

Even though the sample selection in the MaNGA survey was made with a flat $M_{\star}$ distribution, a significant bias exists for high brightness galaxies, i.e. the sample is not complete at higher redshift ranges ($z>0.06$).
Observations out to several $R_e$ would most likely lead to a MOND regime, allowing for a more robust study of modified gravity phenomenology in the higher surface brightness and more massive systems. Furthermore, it has already been shown by Sanders (2000) that in order to reproduce the properties of high surface brightness elliptical galaxies, it is necessary to introduce small deviations from a strictly isothermal and isotropic velocity field in the outer regions.\\

\section*{acknowledgements}

We acknowledge the constructive criticism of an anonymous referee as important towards having reached a clearer and more complete final version.
Reginaldo Durazo acknowledges financial assistance from a CONACyT scholarship and UNAM DGAPA grant IN104517. Xavier Hernandez acknowledges financial assistance from UNAM DGAPA grant IN104517. Bernardo Cervantes Sodi acknowledges financial support through PAPIIT project IA103517 from DGAPA-UNAM. Sebastian F. Sanchez acknowledges financial assistance from UNAM DGAPA grant IA101217. 

Funding for the Sloan Digital Sky Survey IV has been provided by the Alfred P. Sloan Foundation, the U.S. Department of Energy Office of Science, and the Participating Institutions. SDSS-IV acknowledges support and resources from the Center for High-Performance Computing at the University of Utah. The SDSS web site is www.sdss.org.

SDSS-IV is managed by the Astrophysical Research Consortium for the 
Participating Institutions of the SDSS Collaboration including the 
Brazilian Participation Group, the Carnegie Institution for Science, 
Carnegie Mellon University, the Chilean Participation Group, the French Participation Group, Harvard-Smithsonian Center for Astrophysics, 
Instituto de Astrof\'isica de Canarias, The Johns Hopkins University, 
Kavli Institute for the Physics and Mathematics of the Universe (IPMU) / 
University of Tokyo, Lawrence Berkeley National Laboratory, 
Leibniz Institut f\"ur Astrophysik Potsdam (AIP),  
Max-Planck-Institut f\"ur Astronomie (MPIA Heidelberg), 
Max-Planck-Institut f\"ur Astrophysik (MPA Garching), 
Max-Planck-Institut f\"ur Extraterrestrische Physik (MPE), 
National Astronomical Observatories of China, New Mexico State University, 
New York University, University of Notre Dame, 
Observat\'ario Nacional / MCTI, The Ohio State University, 
Pennsylvania State University, Shanghai Astronomical Observatory, 
United Kingdom Participation Group,
Universidad Nacional Aut\'onoma de M\'exico, University of Arizona, 
University of Colorado Boulder, University of Oxford, University of Portsmouth, 
University of Utah, University of Virginia, University of Washington, University of Wisconsin, 
Vanderbilt University, and Yale University.\\

\end{document}